\documentclass{article}
\usepackage{emulateapj,onecolfloat,graphics}

\begin{document}

\twocolumn
[
\title{Completeness in Photometric and Spectroscopic Searches for Clusters}

\author{Martin White${}^1$ and C.S. Kochanek${}^2$}

\affil{
${}^1$Departments of Physics and Astronomy, University of California,
Berkeley, CA 94720\\
${}^2$Harvard-Smithsonian Center for Astrophysics,
60 Garden St., Cambridge, MA 02138}
\affil{email: mwhite@astron.berkeley.edu, ckochanek@cfa.harvard.edu}

\begin{abstract}
\noindent
We investigate, using simulated galaxy catalogues, the completeness of
searches for massive clusters of galaxies in redshift surveys or imaging
surveys with photometric redshift estimates, i.e.~what fraction of
clusters ($M>10^{14}h^{-1}M_\odot$) are found in such surveys.
We demonstrate that the matched filter method provides an efficient and
reliable means of identifying massive clusters even when the redshift
estimates are crude.  In true redshift surveys the method works extremely well.
We demonstrate that it is possible to construct catalogues with high
completeness, low contamination and both varying little with redshift.
\end{abstract}
  ]

\section{Introduction} \label{sec:intro}
 
Clusters of galaxies are one of our most important cosmological probes.
As the most recent objects to form in the universe their number density and
properties are exquisitely sensitive to our modeling assumptions.
Their composition accurately reflects the mix of matter in the universe.
They are bright and can be ``easily'' seen to large distances, allowing
constraints on the crucial interval $0< z\la 1$ where the universal
expansion changes from deceleration to acceleration.
They are located close to their formation site.
Being bright and sparse they are excellent tracers of the large-scale
structure -- they are highly biased so their clustering is easy to measure
and is much more straightforwardly computed from theory than that of
galaxies.

However, constructing large samples of massive clusters for statistical
analyses remains a difficult task.  The original samples
(e.g.~Abell~\cite{Abe58};
 Dalton et al.~\cite{APM};
 Lumsden et al.~\cite{EDCC};
 White et al.~\cite{WBL})
were selected on the basis of projected galaxy overdensity, but it was
quickly realized that such surveys suffer from projection effects and
the large scatter between optical richness and cluster mass
(for recent theoretical studies see
e.g.~van Haarlem, Frenk \& White~\cite{vHaFreWhi};
     Reblinsky \& Bartelmann~\cite{RebBar}).
For this reason attention has broadened to include searches in complete
redshift surveys (e.g. Huchra \& Geller~\cite{HG82};
Geller \& Huchra~\cite{GH83};
Ramella et al.~\cite{Ramella94}; Ramella, Pisani \& Geller~\cite{Ramella97}),
surveys at X-ray wavelengths
(Gioia et al.~\cite{Gio90};
 Edge et al.~\cite{Edg90};
 Henry \& Arnaud~\cite{HenArn};
 Rosati et al.~\cite{Ros95};
 Jones  et al.~\cite{Jon98};
 Ebeling et al.~\cite{Ebe98};
 Vikhlinin et al.~\cite{Vik98};
 Romer et al.~\cite{Rom00};
 Henry~\cite{Hen00};
 Blanchard et al.~\cite{BSBD};
 Scharf et al.~\cite{ROXS}),
and using the Sunyaev-Zel'dovich effect (Carlstrom et al.~\cite{JC}).
More recently there has been significant progress in optical surveys however,
both in terms of data quality and algorithmic sophistication.  The introduction
of accurate, multi-color photometry has allowed estimation of ``photometric
redshifts'' which can mitigate many of the problems of foreground-background
contamination and carefully applied filters can find cluster signals with even
low numbers of cluster galaxies.  

In this work we report preliminary investigations into how well a deep,
multi-color optical survey would find the most massive clusters of galaxies.
We envision this as a first step in a programme which would then obtain
multi-wavelength information about a sample so selected in order to constrain
the evolution of the mass function.
We contrast this with the yield expected from a shallower redshift survey such
as could be done with the Hectospec instrument on the MMT (see Geller~\cite{Geller94}).

\section{Clusters and dark energy}

A recent motivation for revisiting this question, and for investigating
strategies which can allow us to construct a large, well characterized sample
of the rarest clusters over the widest area possible, is the ability of
clusters to shed light on the nature of the dark energy believed to be
causing the accelerated expansion of the universe.
The nature of this dark energy is one of the most vexing problems in cosmology
and one with strong implications for our understanding of fundamental physics.

Since the dark energy is predicted to be smooth, except possibly near the
horizon scale, all of its cosmological effects come in through its effect
on the expansion rate $H(z)$.
Specifically it alters the distance-redshift relations, cosmological volumes
and the growth of perturbations, all of which are integrals of the inverse
Hubble parameter over redshift.
In order to best constrain the dark energy it is desirable to probe the
crucial redshift range $z\simeq 0-1$, where it begins to noticeably
affect the expansion rate, with as much resolution in redshift as possible.
Several authors (most recently Haiman, Mohr \& Holder~\cite{HaiMohHol}) have
suggested using the counts of clusters of galaxies to probe the evolution of
the dark energy in this redshift range.

The strongest cosmological constraints come from relatively massive clusters,
which are intrinsically rare.  In order to construct a large sample of
massive clusters at lower redshifts ($z\la 1$; where detailed followup
observations are conceivable), we need to cover a large area of sky.
This is difficult to do with existing facilities for X-ray or
Sunyaev-Zel'dovich (SZ) observations.
Such a sample, selected optically, would provide a much needed complement to
the higher redshift clusters found by SZ surveys over smaller areas of the
sky.  Once plausible cluster candidates have been found, multi-wavelength
followup is possible (and necessary) to help pin down the `local' sample and
the normalization of the scaling relations which can convert observables into
cluster mass.

\section{Simulated observations} \label{sec:sims}

A realistic search for clusters requires a good match to the spatial
distribution of galaxies and to their mean density, rather than 
a thorough understanding of galaxy formation.  We use high resolution
N-body simulations for the evolution of the dark matter, described
in \S\ref{sec:nbody}, to provide the large scale structure and
clustering of the matter distribution.  We find that N-body based models
are significantly better than Poisson models in describing the fluctuations
in the galaxy background which are important in cluster finding.
Next, we populated the dark matter halos with galaxies as described in
\S\ref{sec:galfake}.
Finally, we produce a simulated observational catalog as described in
\S\ref{sec:gal2map}.
Some of the limitations of our procedure are discussed in
\S\ref{sec:limitations}.

\subsection{N-body simulation} \label{sec:nbody}

On large scales (Mpc and above) the distribution of galaxies will trace that 
of the dark matter, so we can use N-body simulations to provide a model for
the large scale structure and the initial formation of gravitationally bound
halos. We have run a $256^3$ particle simulation of a $\Lambda$CDM model in a
$200h^{-1}$Mpc box using the {\sl TreePM-SPH\/} code
(White et al.~\cite{TreePM}) operating in collisionless (dark matter only)
mode.  This simulation represents a large cosmological volume, to include
a fair sample of rich clusters, while maintaining enough mass resolution
to identify galactic mass halos (see \S\ref{sec:galfake}).
Because it provides a reasonable fit to a wide range of observations,
including the present day abundance of rich clusters of galaxies
(Pierpaoli, Scott \& White~\cite{PieScoWhi}), we have simulated the
``concordance cosmology'' of Ostriker \& Steinhardt~(\cite{OstSte}),
which has $\Omega_{\rm m}=0.3$, $\Omega_\Lambda=0.7$,
$H_0=100\,h\,{\rm km}{\rm s}^{-1}{\rm Mpc}^{-1}$ with $h=0.67$,
$\Omega_{\rm B}=0.04$, $n=1$ and $\sigma_8=0.9$
(corresponding to $\delta_H=5.02\times 10^{-5}$).
The simulation was started at $z=50$ and evolved to the present with the full
phase space distribution dumped every $100h^{-1}$Mpc from $z\simeq 2$ to $z=0$.
The gravitational force softening was of a spline form
(e.g.~Hernquist \& Katz~\cite{HerKat}), with a ``Plummer-equivalent''
softening length of $28\,h^{-1}$kpc comoving.
The particle mass is $4\times 10^{10}h^{-1}M_\odot$ allowing us to find
bound halos with masses several times $10^{11}h^{-1}M_\odot$ and giving
many, many particles in a cluster mass halo ($>10^{14}h^{-1}M_\odot$) to
begin to resolve substructure.

\begin{figure}
\begin{center}
\resizebox{3.5in}{!}{\includegraphics{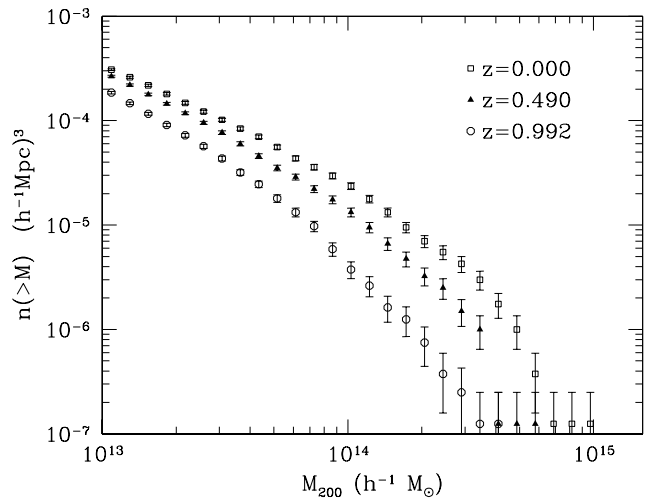}}
\end{center}
\caption{\footnotesize%
The 3D mass function of halos in our simulation box at $z=0$, 0.49 and 0.99.
Masses are $M_{200}$, the mass enclosed within a radius, $r_{200}$, within
which the mean density is 200 times the {\it critical\/} density at that
redshift.  Error bars indicate purely Poisson errors.}
\label{fig:massfn}
\end{figure}

We identify the real clusters in the sample using the 3D dark matter
distribution and the friends-of-friends (FoF) algorithm.
For each cluster we calculate directly from the 3D distribution the mass (we 
use $M_{200}$, the mass enclosed within a radius, $r_{200}$, within which the
mean density is 200 times the {\it critical\/} density at that redshift),
velocity dispersion etc. so we can understand our selection in terms of the
intrinsic, rather than projected, cluster properties.
We define the center of a cluster as the position of the potential minimum,
calculating the potential using only the particles in the FoF group.  This
proved to be more robust than using the center of mass, as the potential
minimum coincided closely with the density maximum for all but the most
disturbed clusters.  
We show the mass function in the box at various redshifts in
Fig.~\ref{fig:massfn}.

\subsection{Adding galaxies} \label{sec:galfake}

We added galaxies to the simulation using a variant of the ``halo model''
for large-scale structure wherein gravitational clustering is described in 
terms of dark matter halos which form a biased tracer of the large-scale 
density field.  Galaxies are distributed in halos following the dark 
matter profile with an occupation number which characterizes the efficiency 
of galaxy formation.
This method produces galaxy distributions which are in agreement with those
produced by semi-analytic models of galaxy formation
(Kauffman et al.~\cite{KCDW};
 Somerville \& Primack~\cite{SomPri};
 Benson et al.~\cite{BCFBL})
and high resolution hydrodynamic simulations including star formation and
feedback
(Katz, Hernquist \& Weinberg~\cite{KatHerWei};
 Gardner et al.~\cite{GKHW};
 Pearce et al.~\cite{Pearce};
 White, Hernquist \& Springel~\cite{WhiHerSpr})
and can match the observed low-order clustering statistics of galaxies
(e.g.~Jing et al.~\cite{JinMoBor};
 Benson et al.~\cite{BCFBL};
 Seljak~\cite{Sel};
 Peacock \& Smith~\cite{PeaSmi};
 Scoccimarro et al.~\cite{SSHJ};
 Scoccimarro \& Sheth~\cite{ScoShe}).

Our methodology is somewhat simpler than the full semi-analytic treatments
described above, more closely approximating that of
Kauffmann, Nusser \& Steinmetz~(\cite{KauNusSte}).
For every output of the simulation we produce a halo catalogue by running a
``friends-of-friends'' (FoF) group finder with a linking length $b=0.2$.
This procedure partitions particles into equivalence classes by linking
  together all particles separated by less than distance $b$.
We keep all groups above 8 particles, which imposes a minimum halo mass of
$3\times 10^{11}h^{-1}M_\odot$.
A slightly smaller minimum mass would be preferable, but with fixed dynamic
range would come at the expense of less volume.
For simplicity we take the halo ``mass'' to be the sum of the particles masses
in the FoF group.
We populate each halo with an integer number, $N$, of ``galaxies''.
Each halo is a host to galaxies of two types.  The first,
 or central galaxy, is placed at the center of mass and inherits the center of
 mass velocity.  Any additional galaxies are assumed to be satellites and are
 laid down tracing the distribution of mass in the halo, including
asymmetry and sub-structure, and inherit the velocity of the nearest dark
matter particle.  This spatial behavior is as seen in a recent hydrodynamic
model of galaxy formation (White, Hernquist \& Springel~\cite{WhiHerSpr})
and is assumed in the halo model.
By having the galaxies trace the 3D density structure in the halo rather than
an azimuthally averaged radial profile
(such as the Navarro, Frenk \& White~\cite{NFW} profile)
we avoid producing artificially ``spherical'' clusters.
For ease of later identification, we tag each galaxy with the mass of its
parent halo and mark ``central'' galaxies as such.

The number of galaxies in each halo is drawn from a distribution whose
moments we take from the semi-analytic models of galaxy formation of
Kauffman et al.~(\cite{KCDW}) as fit by Sheth \& Diaferio~(\cite{SheDia}).
Following Scoccimarro et al.~(\cite{SSHJ}) we model the distribution of
$N$ as a binomial.  For simplicity we use the same $N(M)$ at all redshifts.

Unfortunately, a simple implementation of these algorithms poorly reproduce
real observations because they under predict the observed numbers of galaxies 
by approximately a factor of three.  The missing galaxies arise because the
available models for the halo multiplicity function are calibrated to match 
particular flux-limited samples (e.g. the APM survey) rather than providing
general expressions as a function of galaxy luminosity.

\begin{figure}
\begin{center}
\resizebox{3.5in}{!}{\includegraphics{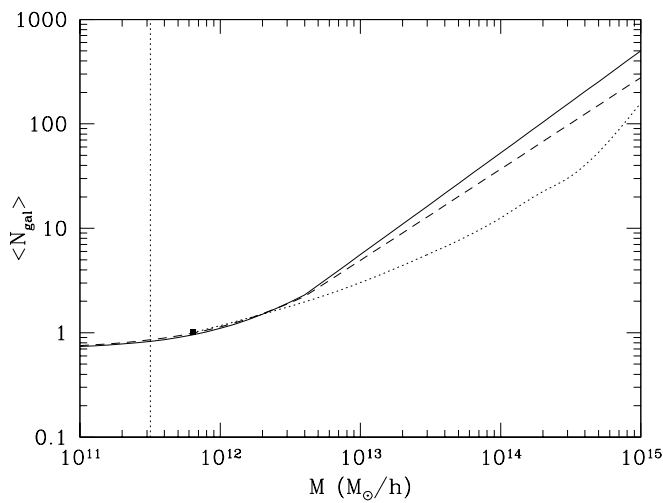}}
\end{center}
\caption{\footnotesize%
The mean number of galaxies in a halo of mass $M$.  The dashed line shows
the fit to the semi-analytic model of Kauffman et al.~(\protect\cite{KCDW}),
the dotted line the estimate of Peacock \& Smith~(\protect\cite{PeaSmi}) which
is zero below $10^{11.8}h^{-1}M_\odot$.
The solid line is the functional form used in this work.
The vertical dotted line marks the mass of an 8 particle halo.}
\label{fig:haloocc}
\end{figure}

Since our ability to characterize the search for clusters depends critically
on the actual numbers of galaxies as well as their spatial distribution, we
adjusted the models to better match the observed density of galaxies.
The basic problem is that the number of galaxies should vary with the minimum
luminosity as $\Gamma[1+\alpha,L/L_*]$ where $\alpha$ is the faint-end slope
of the luminosity function, modeled as a Schechter function (see
Eq.~\ref{eqn:lumfn}) with characteristic luminosity $L_*$.
The standard halo multiplicity expressions were normalized at a luminosity
limit $L/L_* \sim 1/2$ and our model surveys need to include galaxies down
to $L/L_* \sim 1/10$ or even lower to correctly account for the observed
number of galaxies.
We achieve this by steepening the high-mass slope of the multiplicity function
and including galaxies corresponding to lower mass halos.
In clusters we roughly double the number of galaxies in a
$10^{15}h^{-1}M_\odot$ cluster, and with a comparable increase in the number
of galaxies in low mass halos we preserve the contrast between the
clusters and the background.  Because of the limited dynamic range
in the simulations we cannot directly probe smaller mass halos, so we 
considered as `galaxies' a fraction of the un-grouped particles in the 
simulation chosen so as to have about as many ungrouped galaxies as grouped
galaxies.
The ungrouped particles have similar clustering properties to the lowest
  mass halos.

With these modifications our galaxy sample maintains the properties of the
spatial distribution needed for a realistic model while raising the comoving
density of galaxies closer to the observed density.
For example, the galaxy sample has an approximately power-law correlation
function and power spectrum on small scales, over the range
$0.5h^{-1}{\rm Mpc}<r<10h^{-1}{\rm Mpc}$ the galaxy correlation function
is well fit by $\xi(r)=(r/r_0)^{-\gamma}$ with $r_0=5h^{-1}$Mpc and
$\gamma=1.8$ and $\sigma_8^{\rm gal}\simeq 0.9$.  With $\sim 450,000$
galaxies in the $200h^{-1}$Mpc box at $z\simeq 0$, the total comoving density
of galaxies is close to that implied by the LCRS 
(Lin et al.~\cite{Lin96}) luminosity function.  

\subsection{Simulating a field} \label{sec:gal2map}

We simulate an observed field by ``stacking'' different slices through the
box at earlier and earlier output times.  We divide every output up into
6 ``halves'' (top, bottom, left, right, front, back) of
$200\times 200\times 100h^{-1}$Mpc.
A given observational field is then simulated by dividing the line-of-sight up
into $100h^{-1}$Mpc pieces stepping back from the observer.  For each piece we
choose one half of the box at the appropriate redshift, shifted perpendicular
to the line-of-sight by a random amount using the periodicity of the simulation
volume.  A fraction of the galaxies in that half of the box are projected onto
the sky at the appropriate location with the appropriate redshift, including
the peculiar velocity of the galaxy.
We have chosen $100h^{-1}$Mpc as our sampling interval because it is large
enough that edge effects are minimal even for rich clusters while being fine
enough that line-of-sight integrals are well approximated by sums over the
(static) outputs.  However, even though only a small fraction of clusters lie
within $r_{200}$ of a slice boundary, we decided to require that the
orientation and offset change only on every second slice.  Thus if we choose
at one redshift the front of the box the next slice is required to the back.
In this manner a cluster on the boundary is almost always included, though
the periodicity of the box is artificial.

\begin{figure}
\begin{center}
\resizebox{3.0in}{!}{\includegraphics{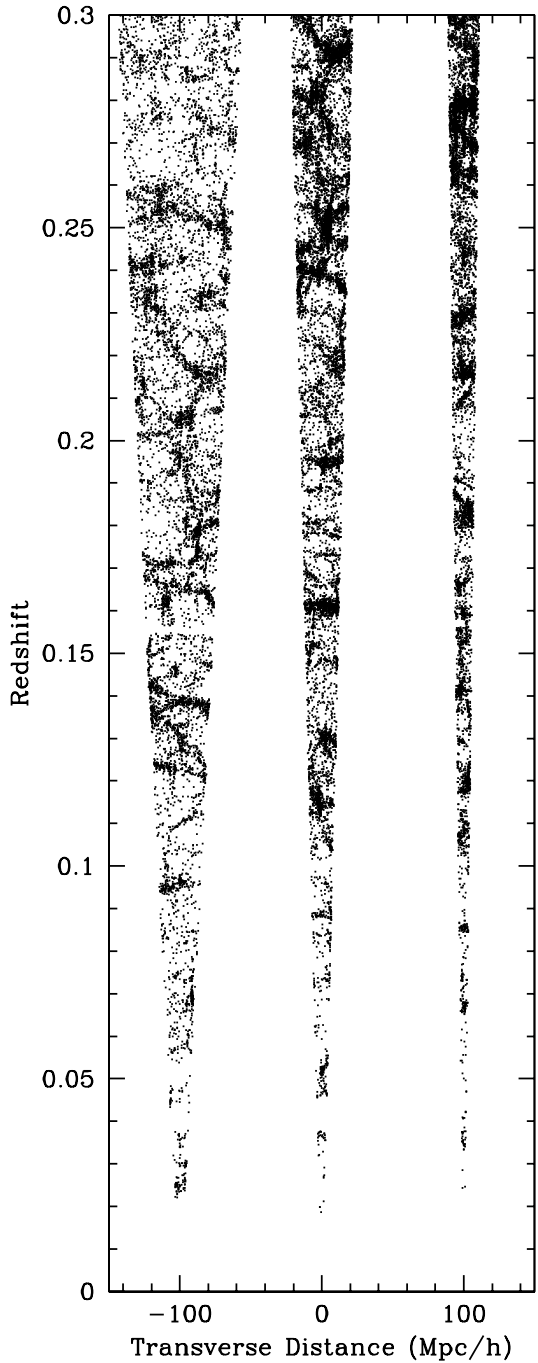}}
\end{center}
\caption{\footnotesize%
Wedge diagrams of parts of 3 of our fields.
In each case we plot redshift against projected separation (comoving) in
right ascension in a wedge $1.5^\circ$ thick.
The surveys, offset for clarity, are our model MMT survey (left) subtending
$6^\circ$ in RA; our model SDSS survey (middle) subtending $3^\circ$ in RA;
and our model LSST survey (right) subtending $1.5^\circ$.
Each dot represents a `galaxy' and all galaxies are  plotted.}
\label{fig:wedge}
\end{figure}

In addition to these fields we also generated ``Poisson'' fields in which the
galaxy positions in each simulation box were randomized before being placed
into the map.  These fields were used to estimate likelihood thresholds for
the cluster finding described in \S\ref{sec:findcl}.  In practice, we found
that the Poisson fields had such low likelihoods for clusters compared to
the real data that they were of little use.

\begin{figure}
\begin{center}
\resizebox{3.0in}{!}{\includegraphics{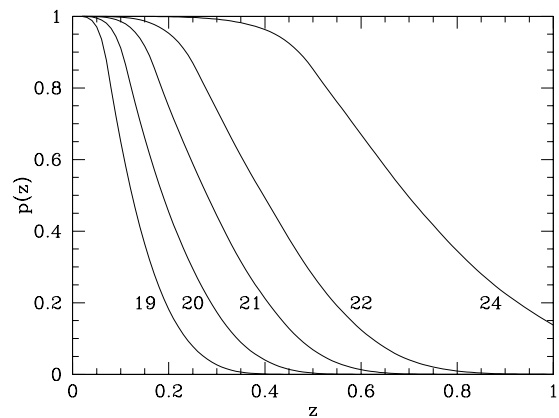}}
\end{center}
\caption{\footnotesize%
The ``selection'' functions adopted for surveys to the limiting
$R$ magnitudes listed.  The lines give the probability $p(z)$ that a 
galaxy at a given redshift is included in the final survey.}
\label{fig:atten}
\end{figure}

Although we assign galaxies to halos based on the mass of their parent halo,
we assign luminosities to the galaxies randomly based on a model luminosity 
function.  The luminosity function enters our calculation only by defining the 
distance-dependent probability that a galaxy is sufficiently luminous to be 
included in the final catalog.  We assume a luminosity function $\phi(L,z)$
where the redshift dependence enters only through the evolution of a 
characteristic luminosity $L_*(z)$ following a conservative $z_f=2$
burst evolution model.  If at redshift $z$ we can detect galaxies brighter
than $L(z)$ given our model flux limit, then the probability of including
a galaxy at redshift $z$ is 
\begin{equation}
    p(z) = \left[ \int_{L(z)}^\infty \phi(L,z) dL \right] 
           \left[ \int_0^\infty \phi(L,z=0) dL \right]^{-1}. 
\end{equation}
For a Schechter luminosity function,
\begin{equation}
  \phi(L) = (n_*/L_*)(L/L_*)^\alpha \exp(-L/L_*),
  \label{eqn:lumfn}
\end{equation}
the function becomes
\begin{equation}
    p(z) = { \Gamma\left[1+\alpha,L(z)/L_*(z)\right] \over 
             \Gamma\left[1+\alpha,0 \right]  }
\end{equation}
where we will base our luminosity function on the LCRS R-band luminosity 
function (Lin et al.~\cite{Lin96}) with $\alpha=-0.70$.  
The resulting number of galaxies is very sensitive to the treatment of the low 
mass, low luminosity galaxies.  Because our simulations do not treat low mass
halos well, we modified the LCRS luminosity function so as to produce 
surveys with galaxy surface densities closer to those observed.  For
luminosities $L_{\rm cut} < 0.1 L_*$ we truncated the luminosity function as
$\phi(L) = (L/L_{\rm cut})\phi_{LCRS}(L_{\rm cut})$.
In a survey to R$=20$~mag,
this modification increases the surface density of galaxies from 
$1100$ per square degree to $1700$ per square degree with no significant
changes in the redshift distribution.  Figure \ref{fig:atten} shows the
selection function $p(z)$ for a range of limiting magnitudes.

For simplicity we do not attempt to assign luminosities or colors to the
galaxies, but characterize them only by the Gaussian uncertainty in their
redshifts.  For typical luminosity functions, the flux of a galaxy is
sufficient to determine the redshift with an uncertainty of $\sigma_z=z/2$.
This sets an upper bound on the redshift uncertainties for nearby galaxies.
For photometric redshifts we will explore $\sigma_z=0.05$ and $0.10$.  For
spectroscopic redshifts we used a very conservative uncertainty of
$\sigma_z=0.01$ to smooth the distribution on scales somewhat larger than
the velocity dispersions of rich clusters.

We considered three survey models motivated by the ongoing or proposed
photometric and redshift surveys.  The first example is a complete redshift 
survey to R$=20$~mag as might be conducted with the Hectospec fiber instrument 
on the 6.5m MMT.  This sample would be ten times deeper (in flux) than the 
current generation of redshift surveys (LCRS, 2dF and SDSS).  
The second example is motivated by the SDSS survey.  It consists of a complete
redshift survey to R$=17.5$~mag, a sparse, red galaxy-biased redshift survey
to R$=20$~mag, and a photometric survey to R$=22$~mag.
The Kauffman et al.~(\cite{KCDW})
model provides separate halo mass-dependent estimates for the number of red
and blue galaxies.  All galaxies brighter than R$=17.5$~mag and 4\% of the
red galaxies (1\% of all galaxies) between R$=17.5$~mag and R$=20$~mag are 
assigned spectroscopic redshifts while the remainder are assigned photometric 
redshifts. The remaining galaxies between R$=17.5$~mag and R$=20$~mag and the 
galaxies between R$=20$~mag and R$=22$~mag are assigned photometric redshifts.
The final example is a deep photometric survey to R$=24$~mag as might 
be done with the LSST.  We assume the survey is conducted in an SDSS region
and includes the SDSS spectroscopic redshifts.  The properties of the
model surveys are summarized in Table~\ref{tab:field}.

\subsection{Limitations} \label{sec:limitations}

The primary limitation in interpreting our results is that our model 
surveys consistently contain too few galaxies. For limiting magnitudes
of R$_c=20$, $22$ and $24$~mag we have $1700$, $8700$ and $32000$ galaxies 
per square degree compared to observed counts of $2400$, $14000$ and $81000$  
galaxies per square degree based on the Gunn-r counts from McLeod \& 
Rieke~(\cite{McLeod95}) and a color of R$_c=r-0.35$.
These undercounts are present despite our modifications to the halo
multiplicity function and the luminosity function.  In the absence 
of numerical resolution effects, simply scaling up $N(M)$ would not affect the
clustering of our galaxies, but may not be the most physically realistic
solution since it implies relatively low mass halos would be hosts to several
galaxies.
While it is plausible that the simulations undercount the halos which will
be low luminosity galaxies or that the luminosity function genuinely turns
down at low luminosity, it would not be physically realistic to raise our
break luminosity above $L_{\rm cut}=L_*/10$. 

However, if our model galaxy distribution adequately reproduces the statistics
of real galaxy distributions, as seems to be the case, the primary consequence
of the lower number of galaxies is to add Poissonian noise to our search.
The Poisson noise level will be 20\%, 27\% and 60\% higher than in a real
survey to R$_c=20$, $22$ and $24$~mag which is not a severe increase.
Since real samples should have more galaxies, our results should be
conservative.

A secondary limitation of our modeling is that we have treated the effects
of evolution in the luminosity function very simply, using a passive
evolution model in which stars form in a single burst at $z_f=2$.
Particularly for the LSST field, where the median
redshift is $z=0.6$, such a model is too simplistic.
A more realistic model would require the identification and treatment of
individual galaxy types.  The inclusion of galaxy types whose evolution is
faster than a passive model would help reduce the discrepancy in the number
of galaxies.  However, given the problems with the halo multiplicity function
and the form of the luminosity function, we felt that adding a more detailed
treatment of evolution should await a better underlying simulation.

\begin{table}
\begin{center}
\begin{tabular}{ccccc}
Name	& $N_{\rm field}$ & Size & $n_{\rm gal}$ & $z_{50}/z_{75}/z_{90}$ \\
MMTS	& 9 & $6.0^\circ\times 6.0^\circ$ & 1700 & 0.23/0.30/0.36 \\
SDSS	& 9 & $3.0^\circ\times 3.0^\circ$ & 8700 & 0.38/0.48/0.57 \\
LSST	& 4 & $1.5^\circ\times 1.5^\circ$ &32600 & 0.59/0.77/0.95
\end{tabular}
\end{center}
\caption{Characteristics of the simulated fields.  Number simulated, size of
field, number density of galaxies (per square degree) and the redshifts
encompassing 50\%, 75\% and 90\% of the survey galaxies.}
\label{tab:field}
\end{table}

\section{Finding clusters} \label{sec:findcl}

Our objective is to automatically produce catalogs of cluster candidates
from the synthetic fields which we can then check using our knowledge of
the true mass distributions.  We do this using the matched filter method
described in \S\ref{sec:mf}, adding some comments on how it can be 
adapted to real data or further improved in \S\ref{sec:improve}.  In
\S\ref{sec:diag} we discuss the diagnostics we use to compare the 
output cluster catalog to the true clusters.  

\subsection{The Matched Filter Algorithm} \label{sec:mf}

We searched for clusters using an automated version of the Adaptive Mesh
Filter (AMF) algorithm (Kepner et al.~\cite{Kepner99}), which is itself
based on the ``matched filter'' algorithm of Postman et al.~(\cite{MF}).

We model the density of galaxies as a redshift-dependent background $\rho_b(z)$
and a distribution of $k=1\cdots n_c$ clusters.  Cluster $k$ is described by
its angular position $\vec{\theta}_k$, (proper) scale length, $r_{ck}$,
redshift $z_k$ and galaxy number $N_k$.
At the corresponding angular diameter distance $D_A(z_k)$ the density of
galaxies associated with the cluster is
\begin{equation}
    N_k \Sigma[(\vec{\theta}-\vec{\theta}_k)D_A(z_k)/r_{ck}]\delta(z-z_k)
\end{equation}
where we use a projected profile
\begin{equation}
  \Sigma(x) \propto {1\over (1+x)^2}
\label{eqn:sigma}
\end{equation}
normalized by $\int_0^{c r_c} \Sigma(x) d^2\vec{\theta} \equiv 1$ for the
angular distribution and
(assuming redshift errors large compared to cluster velocity dispersions)
a delta function for the redshift distribution.  The simple, analytic profile
defined by Eq.~(\ref{eqn:sigma}) provides a good match to a projected NFW 
(Navarro, Frenk \& White~\cite{NFW}) profile.  We fixed the halo concentration
to $c=4$ and the break radius to $r_c=200 h^{-1}$~kpc, as typical parameters
for cluster-mass halos (e.g. Bullock et al.~\cite{Bullock01}). 
For a galaxy $i$ located at $\vec{\theta}_i$ and redshift $z_i$ with assumed
Gaussian uncertainties characterized by $\sigma_i$ the expected density for
cluster $k$ is
\begin{eqnarray}
   \rho_k(\vec{\theta}_i,z_i) =
     N_k \Sigma\left[(\vec{\theta}_i-\vec{\theta}_k)D_A(z_k)/r_{ck}\right]
     \qquad\qquad \nonumber \\
     \qquad\qquad \times
          \exp\left[-(z_i-z_k)^2/2\sigma_i^2\right]/\sqrt{2\pi}\sigma_i
\end{eqnarray}
and the predicted density for galaxy $i$ becomes
\begin{equation}
    \rho_i(\vec{\theta}_i,z_i) = \rho_b(z_i) +
       \sum_{k=1}^{n_c} \rho_k(\vec{\theta}_i,z_i)
\end{equation}   
where the background model must also be modified to include the effects of
the redshift uncertainties.  

The Gaussian redshift uncertainties model any information used to estimate the
redshift of the galaxies in the survey.
At its crudest this estimate comes only from the flux (magnitude) of the
galaxy, and at its best it comes from direct spectroscopic redshifts.
We are interested in the intermediate case where we possess photometric
redshift estimates, presumably derived from galaxy colors, with accuracies
in a range from $0.05 \lesssim \sigma_z \lesssim 0.1$.
 
The likelihood of the model over an area $A$ encompassing all the clusters
and galaxies $i=1 \cdots n_g$ is 
\begin{eqnarray}
   \ln {\cal L} = -A \int dz\ \rho_b(z) - \sum_{k=1}^{n_c} N_k 
     \qquad\qquad \nonumber \\
     \qquad\qquad
 + \sum_{i=1}^{n_g} \ln \left[ \rho_b(z_i) +
   \sum_{k=1}^{n_c} \rho_k(\vec{\theta}_i,z_i) \right].  
\label{eqn:like}
\end{eqnarray}
which is derived from the Poisson statistics of galaxies distributed over
infinitesimal bins in redshift and angle (this is termed the ``fine''
likelihood by Kepner et al.~\cite{Kepner99}, as compared to a ``coarse''
likelihood based on Gaussian statistics).  

We build the model iteratively starting from a smooth background ($n_c=0$).
At each step we use the galaxy positions and redshifts as trial cluster
centers, optimizing the likelihood with respect to the next cluster richness
$N_k$ with $k=n_c$ but holding the properties of the background and all
previous clusters fixed.
We add the trial cluster producing the largest increase in the likelihood to
the global model and then search for the next cluster, continuing the process
until the likelihood gain drops below a threshold.   

Our approach differs from that described by Kepner et al.~(\cite{Kepner99})
in several respects.  First, we make no use of the ``coarse'' (Gaussian)
statistical model.  After careful arrangement of the calculation and
construction of linked lists, our execution time was not dominated by the
optimization of the Poissonian likelihood with respect to $N_k$.
Second, our density model explicitly includes the distribution and structure
of previously found clusters.
This automates the algorithm and provides a reasonable approach to separating
overlapping clusters and minimizing multiple discoveries of the same cluster.
In essence, we have combined the AMF algorithm for finding clusters with the
Clean algorithm of radio astronomy for producing maps.
Third, rather than simply clipping the redshift catalog to bracket the
redshift of a trial cluster, we explicitly include the error-convolved
redshift distribution of the cluster galaxies as part of the density model.

\subsection{Performance, Tuning \& Refinements } \label{sec:improve}

For theoretical convenience we defined our algorithm purely in terms of
redshift uncertainties, although it would be trivial to redefine it in
terms of luminosities and colors.  A combination of the luminosity
function and spectrophotometric models would provide predictions
$m_{\rm est}^j(z)$ for the measured magnitudes $m_i^j$ of galaxy $i$ in
filters $j=1 \cdots n_f$ as a function of redshift and our Gaussian redshift
error is replaced by the fit statistic between the model and the data.
This could include a range of galaxy types and differences in galaxy
properties between the field and clusters. 

We ran our experiments with a fixed cluster scale $r_c=200h^{-1}$~kpc and
concentration $c=4$, although in theoretical models clusters have a range
of scales, $100-500h^{-1}$kpc, and concentrations, $c\simeq 4-8$.
We experimented with varying the scale radius and found that that the
algorithm was biased towards allowing $r_c$ to become unreasonably large
for some, but not all, cluster candidates.
Although we did not conduct further experiments, the problem could be solved
by adding a prior probability term for either the scale radius or the cluster
mass to bias the solutions against finding overly large or massive clusters.
The most natural prior is a simple model for the cluster mass function such
as a $P(N) \propto 1/N^2$ power-law.
We also found that if we increased the outer fit radius too much (i.e.~larger
$c$ at fixed $r_c$) the algorithm systematically merged neighboring clusters.
With $c=4$ this rarely happened, although occasionally a rich, nearby cluster
was split into more than one `candidate'.

We constructed our background model field-by-field based on a coarsely binned
redshift distribution of the galaxies with their assigned redshifts
(i.e.~including the redshift error and scatter).
The continuous distribution was then obtained by linear interpolation between
the bins, which proved sufficient for our purposes and more stable than spline
interpolation.

The initial distribution of likelihoods $\ln{\cal L}$ is relatively well 
modeled as a log-normal distribution with a tail to higher likelihoods.
To set the termination point of our algorithm we fit the initial 
likelihood distribution to determine the mean and dispersion after
rejecting likelihoods more than two standard deviations from the
mean.  We empirically set the stopping point at a likelihood 
threshold corresponding to the mean plus 1.3 standard deviations,
which would mean that 90\% of the galaxies were below the
threshold for a Gaussian distribution.

Finally, we optimized only the properties of the new cluster in estimating the
likelihood.  The performance of the algorithm might be enhanced by
simultaneously optimizing the richnesses of any overlapping clusters.

\subsection{Diagnostics} \label{sec:diag}

After deriving the output cluster catalog we match it to both the input
cluster catalog and the galaxy catalog.  For each galaxy we have the
probability $\rho_b$ that the galaxy is in the field and the probabilities
$\rho_k$ that it is in any of the $k=1\cdots n_c$ cluster candidates.
We assigned galaxies to clusters by first finding the most probable cluster
for the galaxy (the index $k$ with the maximum $\rho_k$ for the galaxy)
and then assigning it to the cluster if $\rho_k > \rho_b$.
For comparison to the fitted cluster number $N_k=N_{\rm fit}$, we also
counted the number of galaxies, $N_\Delta$, above a range of contrast
thresholds where ($\rho_k > \Delta \rho_b$).
Our basic assignment procedure used a contrast $\Delta=1$ and will have
more background contamination than a higher threshold.
We find that $N_1$ tracks $N_{\rm fit}$ closely, but with more scatter.
These estimates for the number of member galaxies can be compared to the
true number, $N_{\rm true}$, of galaxies assigned to cluster.

To match the output cluster catalogue to the input catalogue we used
position and redshift information and in addition the modal parent halo mass
of the galaxies assigned to the output cluster.  This results in a unique
match except in cases where a nearby rich cluster is broken into several
candidates by the group finder, in which case that cluster can be flagged
more than once.
The matching is done in two directions, the best match from the input
catalog to each cluster in the output list and the best match from the
output catalog to each cluster in the input list.  It is the latter,
with duplicates trimmed, that we use to estimate completeness.

It occasionally happens that two clusters overlap on the sky and lie within
$2\sigma$ of each other in the redshift direction.  In these cases our
algorithm often misses one of the clusters, assigning its galaxies to the
overlapping cluster.
For the MMT survey this occurred slightly more than once per field, for a
total of 12 missed clusters in the 9 fields.
As the redshift error increases this number also increases, quadrupling for
$\sigma_z=0.05$.

\section{Results} \label{sec:results}

We illustrate our results by examining our ability to produce a catalog
of clusters with masses above $2\times 10^{14}M_\odot$, as these are the
clusters which provide the greatest constraints for cosmology.
Our assumption is that the catalog is an intermediate step, with further
optical, X-ray or SZ observations being used to clean and calibrate the
sample.  Thus we discuss only the identification of clusters and their
members rather than the derivation of physical parameters.
The selection of the catalog will represent a trade-off between completeness
and contamination, with the contamination arising either from real,
but lower mass, clusters or complete artifacts.  We use our knowledge
of the true cluster properties to design selection procedures for
attaining this goal (see Appendix).

\begin{figure}
\begin{center}
\resizebox{3.5in}{!}{\includegraphics{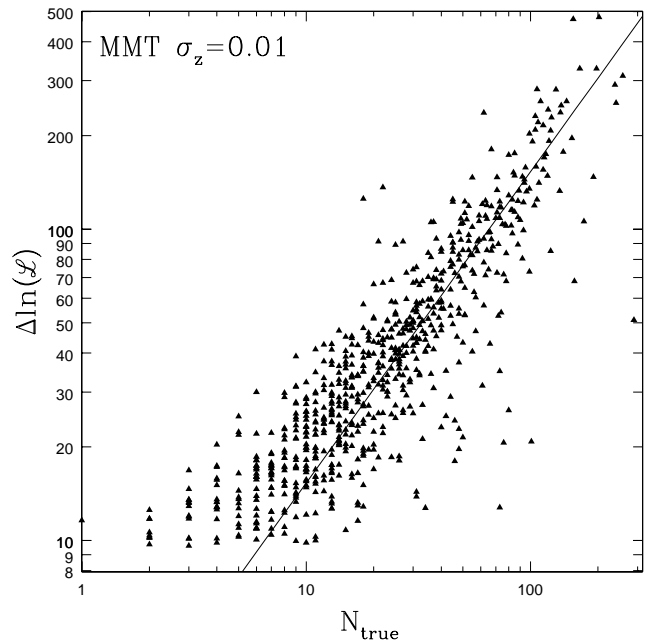}}
\end{center}
\caption{\footnotesize%
Likelihoods for $M \geq 2\times 10^{14}M_\odot$ clusters as a function of
their true galaxy number $N_{\rm true}$.  The line is a linear fit
$\Delta \ln {\cal L}\propto N_{\rm true}$ for the systems with $N_{\rm true}\geq 10$.}
\label{fig:nlike}
\end{figure}

Fig.~\ref{fig:zlike} shows the distribution of matched clusters in likelihood
and redshift for the MMT redshift survey, marking the ones above our 
$2\times 10^{14}M_\odot$ mass threshold.  As expected, higher mass and lower 
redshift lead to higher likelihoods, but there is no sharp boundary
between high and low mass clusters.  However, there is clearly a 
redshift-dependent likelihood threshold which would keep the 
completeness (fraction of $M\geq 2\times 10^{14}M_\odot$ clusters found)
high, the contamination (fraction of $M<2\times 10^{14}M_\odot$ clusters or
false detections) low, and both roughly independent of redshift. If
we simplify the likelihood calculation (Eq.~\ref{eqn:like}) by assuming a top
hat cluster density profile, then we can show that the leading
terms in the likelihood depend only on the number of galaxies in
the cluster, with $\Delta \ln {\cal L} \propto N_{\rm true}$
to lowest order.\footnote{For very large numbers of galaxies the scaling
becomes $\Delta \ln {\cal L} \propto N_{\rm true} \ln N_{\rm true}$.}
As shown in Fig.~\ref{fig:nlike}, the likelihood scales in this manner
for the data as well.
 
\begin{figure*}
\begin{center}
\resizebox{6in}{!}{\includegraphics{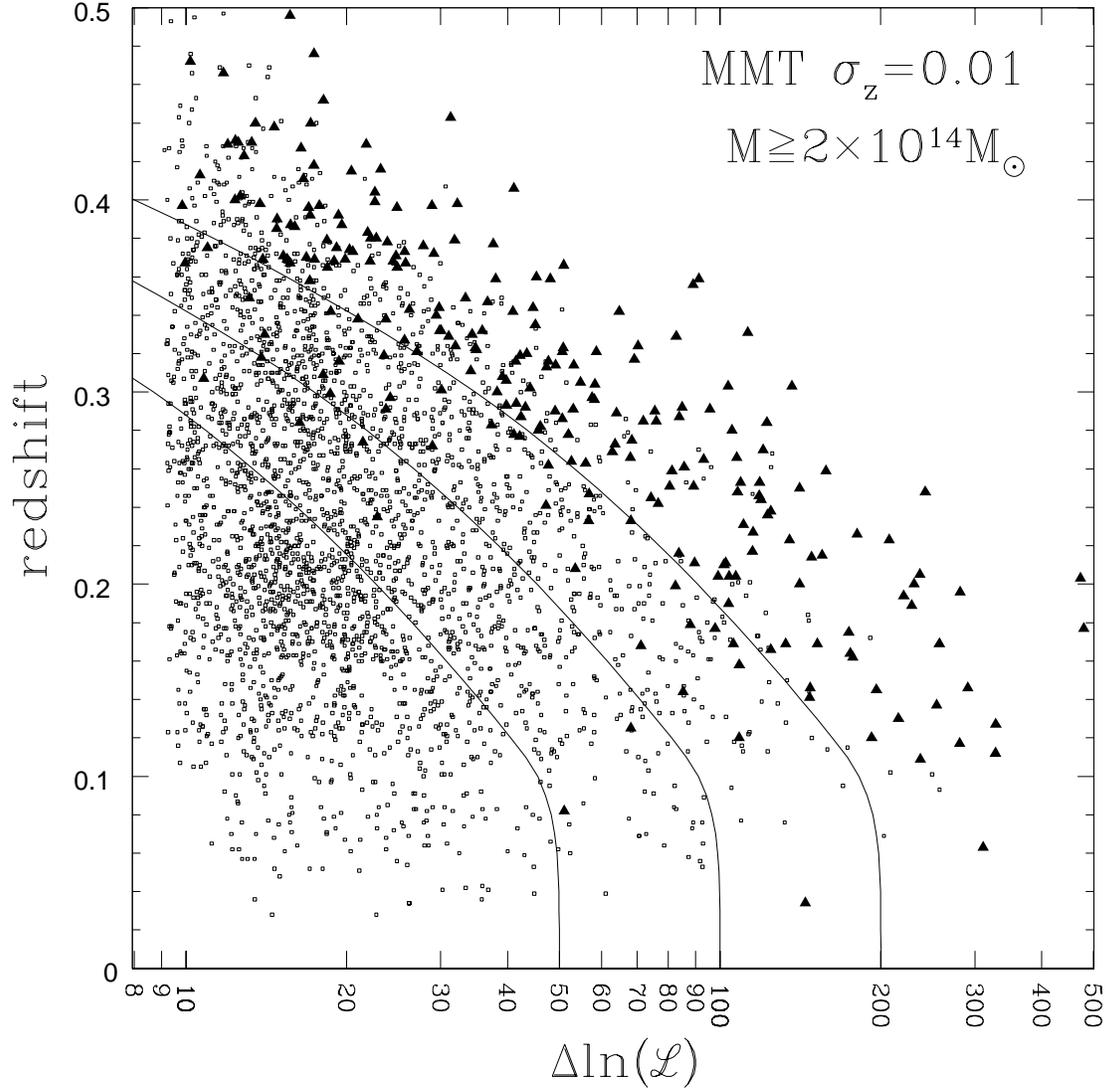}}
\end{center}
\caption{\footnotesize%
The distribution of candidate clusters in likelihood and redshift.
Clusters with masses $M \geq 2\times 10^{14}M_\odot$ are shown by the large,
filled triangles, while those with masses $M<2\times 10^{14}M_\odot$ are
shown by small, open squares.  The curves show cluster selection boundaries
for zero redshift likelihoods of 50, 100 and 200. }
\label{fig:zlike}
\end{figure*}

Thus, although there is considerable scatter due to differences in the
structure of the cluster, the distribution of galaxies and the cluster
environment, we adopt a likelihood threshold designed to track the number 
of galaxies expected in a cluster of fixed mass.  For a Schechter
luminosity function of slope $\alpha$ and an evolving characteristic
luminosity $L_*(z)$, this corresponds to a likelihood threshold which 
decreases as $\Gamma[1+\alpha,L(z)/L_*(z)]$ with redshift, where
$L(z) = 4\pi D_{\rm lum}^2 F$ is the luminosity corresponding to the
survey flux limit.  For further experiments we set our survey 
thresholds using our knowledge of the true masses.  In a real survey
the thresholds would have to be calibrated using clusters of known
mass.  Fig.~\ref{fig:zlike} illustrates the redshift-dependent 
likelihood cuts $\Delta\ln {\cal L}_{\rm cut}(z)$ for a range of local
normalizations $\Delta\ln {\cal L}_{\rm cut}(z=0)$.

We used the most common parent halo mass of the galaxies identified with
a cluster candidate to identify the input halo corresponding to the 
candidate.  This procedure led to multiple identifications of the most
probable, low redshift, massive clusters where we would find lower
likelihood satellite clusters most of whose galaxies are members of
the more massive cluster.  This is at least in part due to our fixed
filter profile whose $r_c=0.2h^{-1}$~kpc is somewhat smaller than the
break radius of the most massive clusters. 
We have automatically filtered these out by dropping
cluster candidates with the same modal mass as a more likely cluster
and within a projected separation of $1h^{-1}$~Mpc and a redshift
difference of $\Delta z=0.05$.  For the MMTS model survey these
represented 5\% of the cluster candidates found.  In a real survey,
where we would lack the knowledge of the parent masses, these 
false candidates would be initially identified as lower mass 
clusters in the halo of a massive cluster and then eliminated 
by more careful modeling of the structure of the most massive
candidates.

We also find real high mass clusters with anomalously low likelihoods.
Many of these are edge effects, where the cluster is within $0.5h^{-1}$~Mpc
of the field edge. We made no modifications to our algorithm to adjust
the likelihoods for the field edges.  We also find a very small number
of overlapping high mass clusters in which one cluster absorbs galaxies
from the other leading to overly high likelihood cluster and one overly
low likelihood cluster.  While the low likelihood cluster may drop 
below our selection thresholds, because it overlaps with a cluster
above the threshold, later studies with more accurate redshifts
will correct the confusion.  Errors in finding the clusters in
the original N-body simulation can also be interpreted as 
completeness and contamination problems -- the FoF algorithm
can combine merging clusters into a single more massive cluster
which our search algorithm then rediscovers as a pair of 
merging clusters of lower likelihood than expected given
the FoF mass estimate.  This effect is somewhat exacerbated by 
our use of the canonical but relatively large linking length $b=0.2$.

\begin{figure*}
\begin{center}
\resizebox{6.0in}{!}{\includegraphics{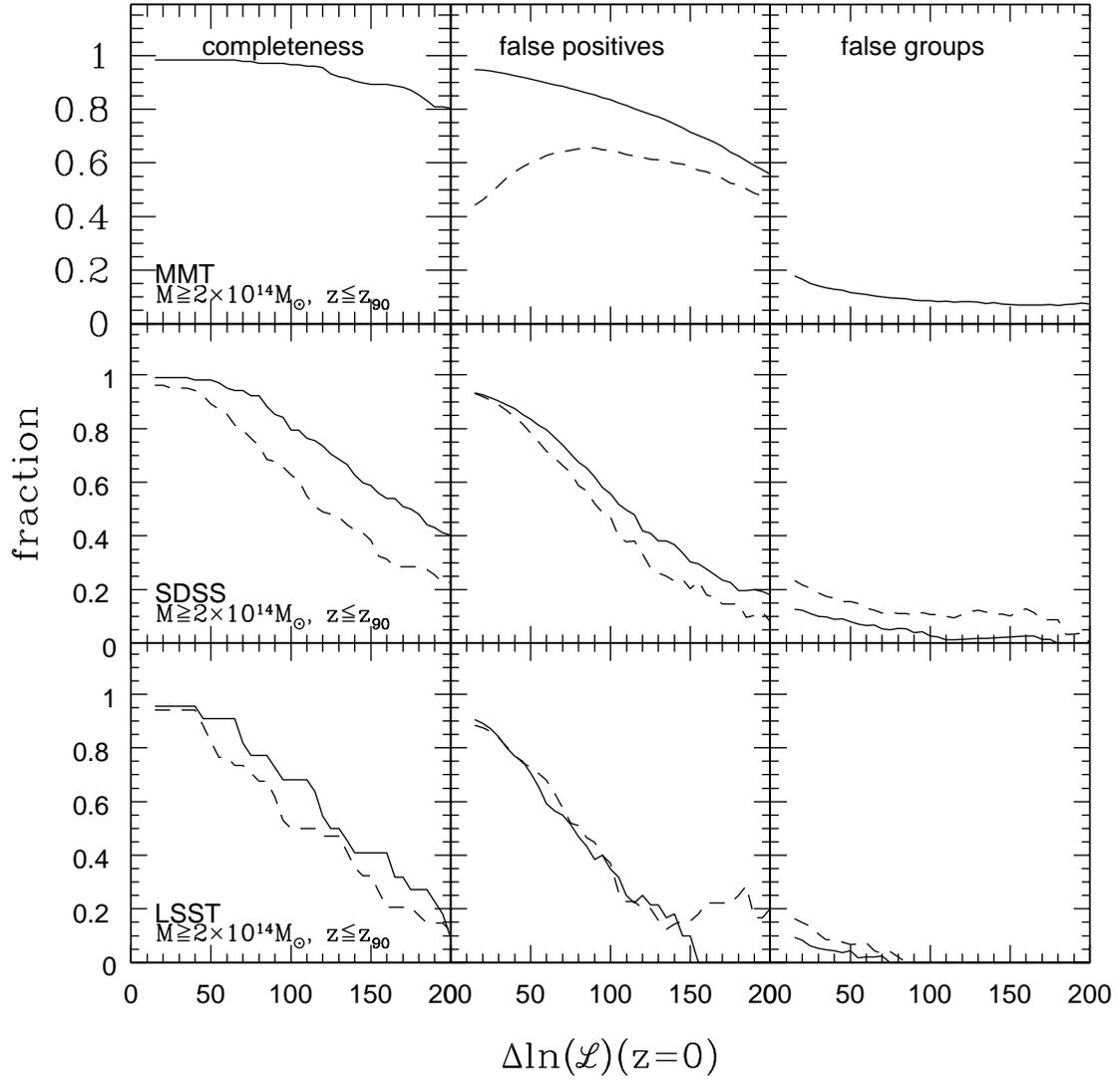}}
\end{center}
\caption{\footnotesize%
The survey completeness and false positive rates.  The top, middle and
bottom rows illustrate the properties of the MMTS, SDSS and LSST model
surveys as a function of the zero redshift likelihood cut 
$\ln{\cal L}(z=0)$.  The left column shows the {\it completeness\/}
defined by the fraction of $M_{200}\geq 2\times 10^{14}h^{-1}M_\odot$ clusters
found in the survey out the redshift encompassing 90\% of the survey galaxies.
The middle column shows the {\it false positive\/} fraction defined by the 
fraction of cluster candidates above the likelihood threshold which are 
not $M_{200}\geq 2\times 10^{14}h^{-1}M_\odot$ clusters.
The right column shows the {\it false group\/} fraction defined by the
fraction of cluster candidates above the likelihood threshold which are
not identified with any input cluster.
For the MMT survey the dashed line in the false positive column shows the
fraction of candidates which correspond to slightly less massive clusters
with $3\times 10^{13}h^{-1}M_\odot\leq M_{200}< 2\times 10^{14}h^{-1}M_\odot$.
For the SDSS and LSST surveys the line patterns show the assumed photometric
redshift errors of $\sigma_z=0.05$ (solid) and $0.10$ (dashed).  }
\label{fig:complete1}
\end{figure*}

\begin{figure*}
\begin{center}
\resizebox{6.0in}{!}{\includegraphics{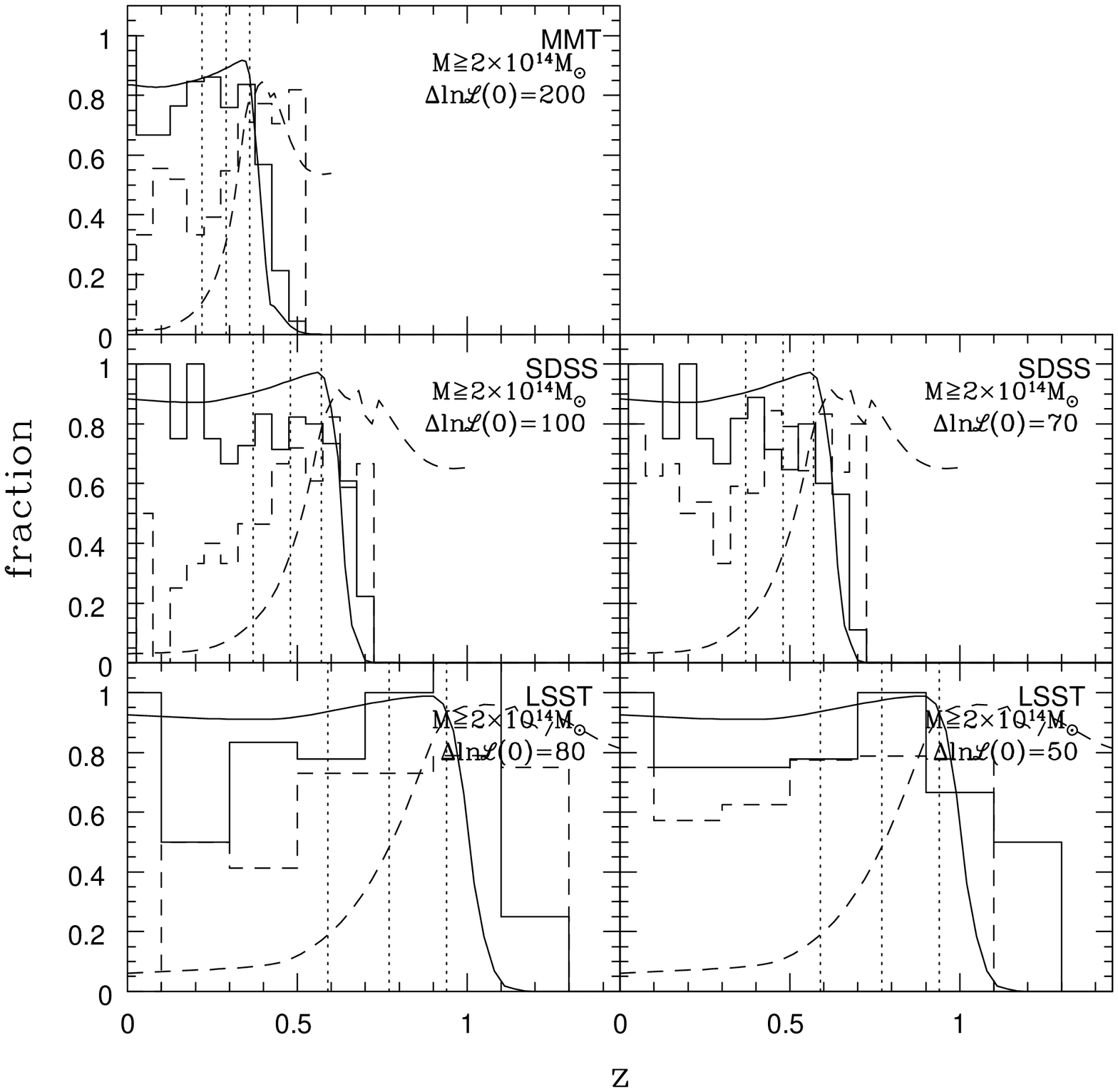}}
\end{center}
\caption{\footnotesize%
The survey completeness and false positive rates.  The top, middle and bottom 
rows illustrate the properties of the MMTS, SDSS and LSST model surveys as a 
function of redshift.  The columns show the effects of increasing errors in
the redshift estimates.  These are fixed to $\sigma_z=0.01$ for the MMT
survey and are $\sigma_z=0.05$ (left) and $0.10$ (right)
for the SDSS and LSST model surveys.  The solid histograms show the 
completeness, the fraction of $M_{200}\geq 2\times 10^{14}h^{-1}M_\odot$
clusters found by the survey, and the dashed histogram show the fraction of
false positives in the survey.  This includes both real, but less
massive clusters and false groups, but is generally dominated by
real clusters with
$3\times 10^{13}h^{-1}M_\odot\leq M_{200}<2\times 10^{14}h^{-1}M_\odot$.
The solid (dashed) curves show the Poisson model for the completeness
(false positive fraction) derived in the Appendix.
The vertical dashed lines mark the redshift encompassing 50\%, 75\%
and 90\% of the survey galaxies. }
\label{fig:complete2}
\end{figure*}

The next step in defining a sample is to select $\Delta\ln{\cal L}(0)$, the
zero redshift normalization of the likelihood selection function.  
Fig.~\ref{fig:complete1} illustrates how the completeness and false
positive fraction depend on the likelihood threshold for each of
our model surveys.  We include all cluster candidates inside the redshift 
encompassing 90\% of the survey galaxies (see Table~\ref{tab:field}).
The equivalent curves for lower redshift thresholds will have lower
false costive rates for the same completeness because the number of
false positives rises with redshift.
We define the completeness as the fraction of clusters inside this redshift
limit with masses above $M_{200}\geq 2\times 10^{14}h^{-1}M_\odot$ which are
found in the cluster catalog with likelihoods above the threshold.
As we raise the likelihood threshold, the completeness declines.
False positives are candidate clusters with likelihoods exceeding the
threshold which do not correspond to a
$M_{200}\geq 2\times 10^{14}h^{-1}M_\odot$ cluster.
We distinguish two types of false positives -- candidates identified with
real but less massive groups, and candidates we could not identify with
any group (a ``false'' group).
Higher likelihood thresholds reduce the numbers of false positives.

Complete redshift surveys, as illustrated by the MMT survey in 
Fig.~\ref{fig:complete1}, easily produce very complete cluster samples to
redshifts well past the survey median.\footnote{Bear in mind that our model
for a redshift survey is very conservative, since a real redshift
survey has velocity measurement errors under 100~km/s and even 
rich clusters have velocity dispersions not much larger than
1000~km/s, while a 1\% redshift error at $z=0.1$ corresponds to
3000~km/s. } While there are few false groups,
the overall false positive fraction is significant and it
is probably impossible to eliminate this problem.  When we combine a
noisy mass estimator (see below) with the very steep cluster mass function
(see Fig.~\ref{fig:massfn}), many apparently massive clusters will be lower
mass clusters with overestimated masses (a form of Malmquist bias).
Most of the false positives in the MMT survey are real groups or clusters
in the mass range $3 \times 10^{13}h^{-1}M_\odot \leq M_{200}\la 10^{14}h^{-1}M_\odot$
(see Fig.~\ref{fig:complete1}).  We discuss this mathematically in
an Appendix.
The completeness of the survey at low redshift is underestimated by our basic matching
software.  Of the 31 clusters missed for low likelihood thresholds, 3 (7) are
within 1 (4) arcmin of a field edge and 4 have virial radii overlapping that
of another massive cluster with a redshift difference smaller than $2\sigma_z$.
As we change the mass threshold, we can maintain high completeness out to
the redshift where the typical cluster at the mass threshold contains three
galaxies, an effect which is well described by the Poisson model for the
survey described in the Appendix.

As we switch from spectroscopic redshifts information to photometric redshifts
the completeness achievable for a given false positive rate declines, as
illustrated in Fig.~\ref{fig:complete1} by the SDSS and LSST survey models.
For runs with larger errors than shown here we even have difficulty
performing the match between the input and output catalogs using the most
common parent halo mass of the galaxies identified with a given cluster
candidate.

Next we selected a likelihood cutoff where we estimate that the survey would
be 80\% complete and determined the completeness and
false positive fractions of the resulting catalogs as a function of redshift
(Fig.~\ref{fig:complete2}).  Because of the design of the likelihood cut
function, the completeness is nearly constant out to the redshift 
encompassing 90\% of the survey galaxies.  The false positive fraction
generally rises with redshift, and catalogs extending to lower redshifts
can have significantly lower contamination for the same level of 
completeness.
The apparent drop in the completeness of the MMT model survey at low redshift
is due in part to edge effects but also to the fact that we have required an
extremely high likelihood threshold to remove low mass systems.
The false positive rate rises faster with redshift in the SDSS survey because
complete redshift information is available only for the nearby galaxies
($z <0.2$).  At these low redshifts, the addition of the deeper photometric
catalog to the redshift data appears to improve the performance
significantly.  

The redshift dependence of the completeness is well defined by the Poisson
model for the survey developed in the Appendix.  For a likelihood threshold
roughly corresponding to the number of galaxies in a cluster at the threshold 
mass, the survey will be nearly complete up to the redshift where the average
number of galaxies in the threshold mass cluster drops below about 3 galaxies.
The Poisson model works less well for explaining the fraction of false 
positives.  Adding small number of background galaxies to each cluster 
does explain the rapid rise in the false positive fraction with redshift.
The model underpredicts the false positive fraction at lower redshifts,
probably because the contamination from the background distribution of 
galaxies is poorly described by a simple Poisson model.  

\begin{figure*}
\begin{center}
\resizebox{6in}{!}{\includegraphics{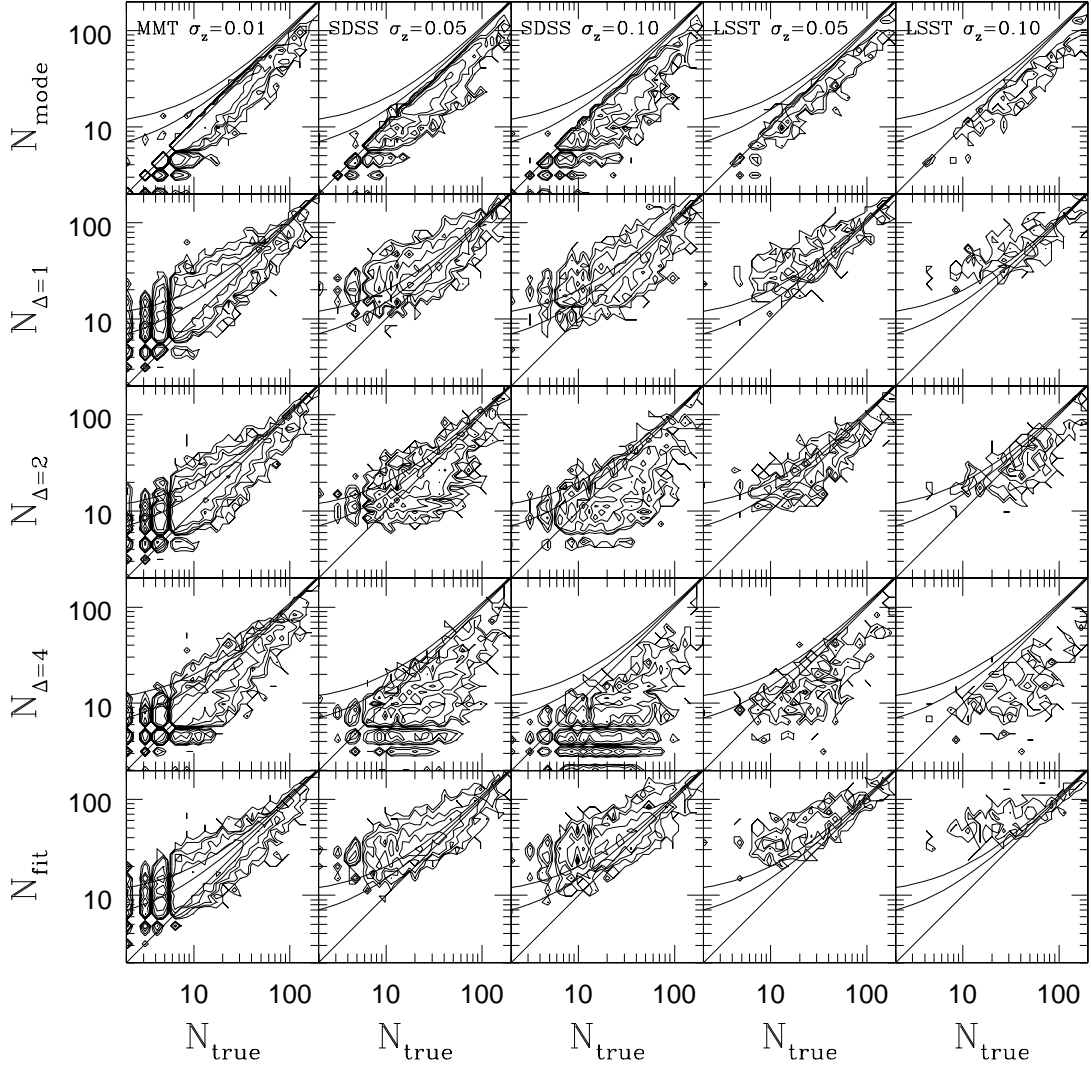}}
\end{center}
\caption{\footnotesize%
Density contours for various estimates of the number of galaxies in the cluster
versus the true number of galaxies $N_{\rm true}$. From bottom to top we 
compare the true number of galaxies to the estimated number from the
likelihood function $N_{\rm fit}=N_c$, the number of galaxies 
$N_{\Delta=4}$, $N_{\Delta=2}$ and $N_{\Delta=1}$ with a likelihood contrast relative to the
background larger than a factor of $\Delta=4$, $2$ and $1$ respectively, and
the number of real cluster galaxies with a likelihood contrast
relative to background larger than unity $N_{\rm mode}$. The contours
are spaced by factors of 2 in the density.  The smooth curves show lines 
where $N=N_{\rm true}$, $N=N_{\rm true}+5$ and $N=N_{\rm true}+10$.}
\label{fig:number}
\end{figure*}

Finally, we explore the correlation of our model cluster parameters with 
the true properties of the cluster.  Fig.~\ref{fig:number} compares 
estimates for the number of cluster galaxies with the true number.  
The number of galaxies estimated in the likelihood, $N_c$, closely matches
the true number in rich clusters.  But, in a catalog selected based on
the cluster likelihood, we tend to find clusters with small galaxy
excesses compared to the real cluster.  Roughly speaking, the fit 
parameter $N_c$ usually finds 5--10 more galaxies than were actually
in the cluster.  We can also estimate the number of galaxies by 
counting the number of galaxies $N_\Delta$ whose probability of
being a cluster member exceeds their probability of being in the
background by a factor of $\Delta$.  The number at unit contrast,
$N_1$, tracks $N_{\rm fit}$ closely with some additional scatter.  Higher
contrast values provide better discrimination against background
contamination.  Fig.~\ref{fig:number} also compares $N_2$ and $N_4$ 
with $N_{\rm true}$.  At an intermediate contrast $\Delta=2$, $N_2$
is less biased systems with small numbers of galaxies but begins
to underestimate the numbers of galaxies in systems with large
numbers of galaxies.  These trends become clearer for the higher
contrast level of $\Delta=4$.  Inspection of
individual systems with extreme ratios $N_{\rm fit}/N_{\rm true}$ provides no
guidance towards an improved estimator.  They tend to be relatively
massive systems with modest numbers of galaxies where the finger-of-god
from the velocity dispersion overlaps a larger than average number
of galaxies.

\begin{figure}
\begin{center}
\resizebox{3.5in}{!}{\includegraphics{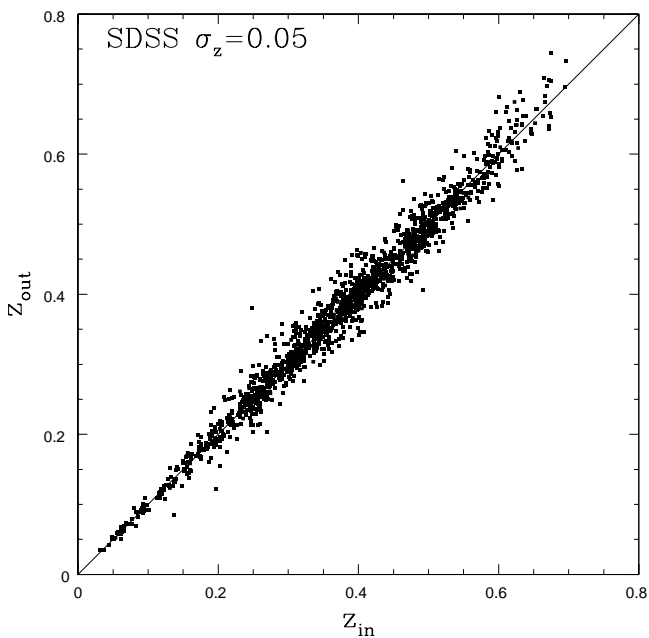}}
\end{center}
\caption{\footnotesize%
Estimated versus true cluster redshifts.}
\label{fig:zdiff}
\end{figure}

Finally we can use the probability that any galaxy is a cluster member to
estimate the cluster redshift.  The estimates scale as expected, as shown
in Fig.~\ref{fig:zdiff}.  In general it is possible to compute any cluster
property (e.g.~velocity dispersion) using such a probability weighting.
We shall defer discussion of such estimators to a future publication.

\section{Discussion} \label{sec:discussion}

Our tests of the matched filter method for finding clusters in optical surveys
with either photometric or spectroscopic redshifts show that it is an efficient
and reliable means of identifying massive clusters even when the redshift 
estimates are crude.  In redshift surveys, where cluster surveys have usually
used FoF methods rather than matched filters, the method works extremely well.
By selecting clusters using a redshift-dependent likelihood threshold roughly
tracking the expected number of galaxies from a cluster of fixed mass, we can
construct catalogs with high completeness, low contamination and both varying
little with redshift.
The method automatically assigns a probability that each galaxy is a member
of any cluster, which can be used in the estimate cluster properties such as
redshift or velocity dispersion.

Both the completeness and the contamination in our mock surveys can be
understood in part using a simple analytic model (described in the Appendix).  The
largest effect is the well known scatter in optical richness at a fixed
cluster mass.  Due to the steeply falling mass function of clusters this
scatter implies that any sample selected on the basis of a fixed number of
galaxies will be contaminated by abnormally rich, low-mass clusters.
In our mock surveys this is by far the largest effect, with ``false''
clusters being almost entirely absent.
We find that the false positive rate increases with redshift, so that our
samples will have less contamination if restricted to a lower redshift
cutoff than the 90th percentile we have assumed throughout.
In any case, the likelihood threshold can be adjusted to modify either the
desired completeness or contamination.  

The addition of photometric data for fainter galaxies to a redshift survey,
as in the SDSS at low redshift, considerably improves the detection of 
clusters over the redshift data alone.  The redshift catalog,
by pinning down the foreground contamination, probably improves the
detection of higher redshift clusters which are detectable only in
the photometric data.  

We thus expect that the matched filter method can in future be used to
construct large samples of clusters to ``modest'' redshifts, though follow-up
observations will be necessary to clean the sample.
Apart from clusters lying near the edge of our fields, the most common
misidentification was to split very large clusters into a core and satellite
population.  This occurred due to our assumption of fixed core radius and
concentration.  The second most common misidentification was when two massive
clusters overlapped, with one `stealing' galaxies from another and causing it
to drop below our likelihood threshold.
In both cases any confusion would be immediately eliminated by more careful
follow-up observations and modeling of the cluster region.

Our investigation is but a first step, and further work is needed.
The primary limitation of our mock surveys is the inadequacy of current
methods for populating dark matter halos with galaxies.  Our simulations
have too few galaxies compared to real surveys (by factors 30\%,
40\% and 60\% for R$<20$, $22$ and $24$~mag).
While simulations with a higher dynamic range would be a step in the right
direction, significant uncertainties remain in the modeling of $N(M)$ and
how it depends on luminosity and redshift.
Encouragingly our results should be conservative in this respect.
It is also necessary to apply this method to real data, to uncover any
failure modes which have been missed by the simulations.

\section*{Acknowledgments}
 
We would like to thank Jeremy Kepner for comments on the manuscript.
C.S.K. was supported by the Smithsonian Institution.
M.W. was supported by a Sloan Fellowship and the National
Science Foundation.  Simulations were carried out at CPAC, through
grants PHY-9507695.

\appendix

\section{Poisson Theory Of Completeness}

The likelihood of finding a cluster is largely controlled by the number of 
member galaxies.  This allows us to make a model for the tradeoff between
completeness and contamination.
If the expected number of galaxies in a cluster of mass $M$ and redshift $z$
is $\langle N \rangle = N_0(M) p(z)$ (see \S3.3) and the halo mass function is $dn/dM$
then the number of clusters with $N_{\rm obs}$ galaxies is
\begin{equation}
    { d n \over d N_{\rm obs} } = \int dV \int dM { d n \over d M }
       { \langle N (M,z) \rangle^{N_{\rm obs}} \over N_{\rm obs}! }
         \exp(-\langle N (M,z) \rangle)
\end{equation}
for volume element $dV$ and assuming Poisson statistics, expected for the
high mass end of the mass function.
If we search for clusters of mass $M \geq M_0$, the total number within
redshift $z$ is
\begin{equation}
   N_{\rm tot}(\geq M_0) = \int_0^z dV \int_{M_0}^\infty dM { dn \over dM }.
\end{equation}
The cluster likelihood roughly scales as $\Delta \ln {\cal L} \propto N$ so we
will find clusters above a fixed mass threshold if we scale the likelihood or
the number of members with the expectation value.
The threshold is set by the number of galaxies at $z=0$, $n_0$, and then
decreases with redshift as $n(z)=n_0p(z)$.
The cluster sample will contain
\begin{equation}
   N_{\rm fnd}(\geq M_0,\geq n_0) =
   \int_0^z dV \int_{M_0}^\infty dM { dn \over dM }
    P\left[n(z),\langle N (M,z) \rangle\right]
\end{equation}
galaxies above the mass threshold, where
\begin{equation}
     P[n,N] \simeq { \Gamma[1+n,N] \over \Gamma[1+n] }
\end{equation}
is the fraction of clusters expected to have $N$ galaxies containing at least
$n$ galaxies.
The completeness of a sample selected with this criterion is
$N_{\rm fnd}/N_{\rm tot}$.
Since the likelihood depends only on the number of galaxies, we also find false
positives from lower mass clusters with galaxy membership above the threshold.
The number of false positives is
\begin{equation}
   N_{\rm false}(< M_0,\geq n_0) =
   \int_0^z dV \int_0^{M_0} dM { dn \over dM }
    P\left[n(z),\langle N (M,z) \rangle\right] ,
\end{equation}
and the fraction of cluster candidates where are false positives is
$N_{\rm false}/(N_{\rm fnd}+N_{\rm false})$.

We can extend this basic theory to a more realistic model for a cluster
survey with two modifications. First,  a cluster must contain a minimum
number of galaxies, $N_{thresh}\simeq 3$, to be detected.  This lower
bound corresponds to the likelihood threshold of the catalog, below
which the candidates are dominated by true false positives with no
correspondence to any cluster.  We implement it in the Poisson model
by using a threshold $n(z) =\hbox{max}(n_0 p(z),N_{thresh})$. Second,
the cluster catalogs are also contaminated by unrelated galaxies.
These chance projections alter the apparent number of galaxies
associated with a cluster.  Based on Fig.~\ref{fig:number} we model
these chance projections as a Poisson process with an expectation
value of $N_b\simeq 5$.  For a cluster expected to have
$N_c$ galaxies and a detection threshold of $i$, the probability
of the cluster including chance projections having at least
$n$ galaxies at least $j$ of which are cluster members is
\begin{equation}
   P(n,j|N_c,N_b) = 
        \sum_{i=0}^{n-j} P(n-i|N_c) { N_b^i \over i! } \exp(-N_b).
\end{equation} 
We are taking the sum $i$ over the possible level of background
contamination, weighted by its Poisson likelihood given the value
of $N_b$, multiplied by the probability that the cluster will
contain enough galaxies for the sum of the number in the cluster
and in the background to reach the threshold.  

We illustrate the behavior of this model in Fig.~\ref{fig:complete2}.
We fixed $N_{thresh}=3$ and we adjusted $n(0)$ to produce a low
redshift completeness slightly above the average observed 
completeness.  We scaled it to be slightly above because we
lose some clusters due to effects not in the model (edges,
overlapping clusters).  We fixed the amount of background
contamination to $N_b=5$, $10$ and $15$ for the MMT
SDSS and LSST model surveys based very crudely on the offset
between $N_{fit}$ and $N_{true}$ in Fig.~\ref{fig:number}. 
The Poisson model describes the completeness of the survey
well, matching the extended region of nearly constant 
completeness followed by a sharp drop produced by the
requirement for a finite number of galaxies $N_{thresh}$ in
a cluster.  The model describes the false positive fraction
less well.  The rapid rise in the false positive fraction
near the drop in the completeness is due to the Poisson
fluctuations in the contamination.  However, the overall
distribution of false positives cannot be explained by 
the Poisson model.  

The limitation of the Poisson model is implicit in the wide
range of likelihoods found for a fixed true number of galaxies 
(see Fig.~\ref{fig:nlike}).  While the likelihood roughly
scales with the true number of galaxies in the cluster, there
is significant scatter about the general trend.  Clusters
differ not only in their total galaxy content, but also in
their internal structure (break radius, concentration), the
sampling of the internal structure, and the density of their
local environment.  Any effects which increase the scatter
between the likelihood and $N_{true}$ will produce more
false positives for a fixed level of completeness.  The 
background contamination is also more complicated than a
the simple Poisson model, since the background galaxies
are themselves clustered.  For example if the average
background contamination is $N_b=4$ but galaxies are 
always clustered in pairs, the likelihood of 6 
contaminating galaxies is nearly doubled.


\begin{thebibliography}{99}
\bibitem[1958]{Abe58}
  Abell G.O., 1958, \apjs, 3, 211
\bibitem[2000]{BCFBL}
  Benson A.J., et al., 2000, \mnras, 311, 793
\bibitem[2000]{BSBD}
  Blanchard A., Sadat R., Bartlett J.G., le Dour M., 2000, \aa, 362, 809
\bibitem[2001]{Bullock01}
  Bullock, J.S., Kolatt, T.S., Sigad, Y., 
    Somerville, R.S., Kravtsov, A.V., 
    Klypin, A.A., Primack, J.R., \& Dekel, A., 2001, 
    MNRAS, 321, 559
\bibitem[2000]{JC}
  Carlstrom J.E., et al., 2000, Phys. Scripta, 85, 148
\bibitem[1996]{Dal96}
  Dalcanton J., 1996, \apj, 466, 92
\bibitem[1992]{APM}
  Dalton G.B., Efstathiou G., Maddox S.J., Sutherland W.J., 1992, \apj, 390, L1
\bibitem[1998]{Ebe98}
  Ebeling H., et al., 1998, \mnras, 301, 881
\bibitem[1990]{Edg90}
  Edge A., Stewart G.C., Fabian A.C., Arnaud K.A., 1990, \mnras, 245, 559
\bibitem[2001]{GKHW}
  Gardner, J.P., Katz, N., Hernquist, L., Weinberg, D.H., 2001, ApJ, 559, 131
  [astro-ph/9911343]
\bibitem[1983]{GH83}
  Geller, M.J., \& Huchra, J.P., 1983, ApJS, 52, 61
\bibitem[1994]{Geller94}
  Geller, M.J., 1994, RAS Canada, 88, 283 
\bibitem[1990]{Gio90}
  Gioia I.M., et al., 1990, \apjs, 72, 567
\bibitem[2001]{HaiMohHol}
  Haiman Z., Mohr J., Holder G., 2001, \apj, 553, 545
\bibitem[1991]{HenArn}
  Henry J.P., Arnaud K.A., 1991, \apj, 372, 410
\bibitem[2000]{Hen00}
  Henry J.P., 2000, \apj, 534, 565
\bibitem[1989]{HerKat}
  Hernquist, L. \& Katz, N., 1989, ApJS, 70, 419
\bibitem[1982]{HG82}
  Huchra, J.P., \& Geller, M.J., 1983, ApJ, 265, 356
\bibitem[1998]{JinMoBor}
  Jing Y.P., Mo H.J., Borner G., 1998, \apj, 494, 1
\bibitem[1998]{Jon98}
  Jones L.R., et al., 1998, \apj, 495, 100
\bibitem[1999]{KCDW}
  Kauffmann G., Colberg J.M., Diaferio A., White S.D.M., 1999, \mnras, 303, 188
\bibitem[1997]{KauNusSte}
  Kauffmann G., Nusser A., Steinmetz M., 1997, \mnras, 286, 795
\bibitem[1999]{KatHerWei}
  Katz N., Hernquist L., Weinberg D.H., 1999, ApJ, 523, 463
\bibitem[1999]{Kepner99}
  Kepner, J., Fan, X., Bahcall, N., Gunn, J., Lupton, R., \& Xu, G., 1999,
    \apj, 517, 78 
\bibitem[1996]{Lin96}
  Lin, H., Kirshner, R.P., Shectman, S.A.,
  Landy, S.D., Oemler, A., Tucker, D.L., \&
  Schechter, P.L., 1996, ApJ, 464, 60
\bibitem[1996]{EDCC}
  Lumsden S.L., Nichol R.C., Collins C.A., Guzzo L., 1992, \mnras, 258, 1
\bibitem[1995]{McLeod95}
  McLeod, B.A., \& Rieke, M.J., 1995, ApJ, 454, 611
\bibitem[1996]{NFW}
  Navarro J., Frenk C.S., White S.D.M., 1996, ApJ, 462, 563
\bibitem[1995]{OstSte}
  Ostriker J., Steinhart P.J., 1995, Nature, 377, 600
\bibitem[2000]{Pea}
  Peacock J.A., 2000, MNRAS, 318, 1144 [astro-ph/0002013]
\bibitem[1996]{PD96}
  Peacock J.A., Dodds S.J., 1996, \mnras, 280, L19
\bibitem[2000]{PeaSmi}
  Peacock J.A., Smith R.E., 2000, preprint [astro-ph/0005010]
\bibitem[1999]{Pearce}
  Pearce, F.R. et al., 1999, ApJ, 521, 99
\bibitem[2001]{PieScoWhi}
  Pierpaoli E., Scott D., White M., 2001, \mnras, 325, 77
\bibitem[1996]{MF}
  Postman M., et al., 1996, AJ, 111, 615
\bibitem[1974]{PS}
  Press W.H., Schechter P., 1974, ApJ, 187, 452
\bibitem[1994]{Ramella94}
  Ramella, M., Diaferio, A., Geller, M.J., \& Huchra, J.P., 1994, AJ, 107, 1623
\bibitem[1997]{Ramella97}
  Ramella, M., Pisani, A., \& Geller, M.J., 1997, AJ, 113, 483
\bibitem[1999]{RebBar}
  Reblinsky K., Bartelmann M., 1999, A\&A, 345, 1
\bibitem[2000]{Rom00}
  Romer A.K., et al., 2000, \apjs, 126, 209
\bibitem[1995]{Ros95}
  Rosatti P., Della Ceca R., Burg R., Norman C., Giacconi R., 1995,
    \apj, 445, L11
\bibitem[2000]{ROXS}
  Scharf C.A., et al., 2000, \apj, 528, L73
\bibitem[2001]{ScoShe}
  Scoccimarro R., Sheth R., 2001, preprint
\bibitem[2001]{SSHJ}
  Scoccimarro R., Sheth R., Hui L., \& Jain B., 2001, ApJ, 546, 20
\bibitem[2000a]{Sel}
  Seljak U., 2000, MNRAS, 318, 203 [astro-ph/0001493]
\bibitem[2001]{SheDia}
  Sheth R., Diaferio A., 2001, MNRAS, 322, 901
\bibitem[1999]{SomPri}
  Somerville R., Primack J., 1999, \mnras, 310, 1087
\bibitem[1997]{vHaFreWhi}
  van Haarlem M.P., Frenk C.S., White S.D.M., 1997, \mnras, 287, 817
\bibitem[1998]{Vik98}
  Vikhlinin A., et al., 1998, \apj, 503, 77
\bibitem[2001]{HaloMass}
  White M., 2001, A\&A, 367, 27 [astro-ph/0011495]
\bibitem[2001]{WhiHerSpr}
  White M., Hernquist L., Springel V., 2001, ApJ, 550, L129
\bibitem[2001]{TreePM}
  White M., Springel V., Hernquist L., 2001, in preparation.
\bibitem[1999]{WBL}
  White R., et al., 1999, \aj, 118, 2014
\end{thebibliography}
\end{document}